\documentclass[letter,
               ]{jacow}
\usepackage{pdfpages,multirow,ragged2e} %
\makeatletter%
	\ifboolexpr{bool{xetex}}
	 {\renewcommand{\Gin@extensions}{.pdf,%
	                    .png,.jpg,.bmp,.pict,.tif,.psd,.mac,.sga,.tga,.gif,%
	                    .eps,.ps,%
	                    }}{}
\makeatother

\ifboolexpr{bool{xetex} or bool{luatex}} 
 {}                                      
 {\usepackage[utf8]{inputenc}}           

\usepackage[USenglish]{babel}
\usepackage{amsmath}
\ifboolexpr{bool{jacowbiblatex}}%
 {%
  \addbibresource{jacow-test.bib}
  \addbibresource{biblatex-examples.bib}
 }{}
\begin{document}

\title{High Precision RF Pulse Shaping with Direct RF Sampling for Future Linear Accelerators
\thanks{This work was supported by the U.S. DOE, Office of Science contract DE-AC02-76SF00515.}}

\author{C. Liu\thanks{chaoliu@slac.stanford.edu}, A. Dhar,  R. Herbst, E. A. Nanni \\ SLAC National Accelerator Laboratory, Menlo Park, California, USA \\
		}
	
\maketitle

\begin{abstract}
In various of particle accelerator designs, amplitude and phase modulation methods are commonly applied to shape the RF pulses for implementing pulse compressors or compensating for the fluctuations introduced by the high-power RF components and beam loading effects. Phase modulations are typically implemented with additional phase shifters that require drive or control electronics. With our recent next-generation LLRF (NG-LLRF) platform developed based on direct RF sampling technology of RF system-on-chip (RFSoC) devices, RF pulse shaping can be realized without the analogue phase shifters, which can significantly simplify the system architecture. We performed a range of high-power experiments in the C-band to evaluate the RF pulse-shaping capabilities of the NG-LLRF system at different stages of the RF circuits. In this paper, the high-power characterization results with the Cool Copper Collider (C\(^3\)) structure driven by RF pulses with different modulation schemes will be described. With the pulse modulation and demodulation completely implemented in the digital domain, the RF pulse shaping schemes can be rapidly adapted for X-band structures simply by adding analogue mixers. 
\end{abstract}

\section{Introduction}
The low-level RF (LLRF) systems of particle accelerators are typically implemented to minimize the field fluctuation in the accelerating structures. For accelerators operating with a single bunch per pulse, the pulse-to-pulse fluctuation control is adequate. However, for accelerators operating with multiple bunches per RF pulse, fluctuation within the pulse becomes critical. Therefore, RF pulses are required to be shaped to compensate for the fluctuation introduced by the RF components or generated by beam loading. The LLRF systems are also used to control other parts of other RF stations of accelerators. For instance, a SLAC Energy Doubler (SLED) based pulse compressor, which requires phase reversal, can be realized by shaping the RF pulse in the LLRF system. For a conventional LLRF system, there is limited flexibility in shaping the RF pulse. Phase reversal for the SLED or other pulse-shaping modulation was implemented with separate RF modules \cite{sled1,sled2}. During the past several years, the SLAC team has designed and implemented the next-generation LLRF (NG-LLRF) system based on RF system-on-chip (RFSoC) technology for linear accelerators (LINACs) \cite{liu2024next}. 

The RFSoC integrates RF ADCs that can sample RF signal with frequency up to 6 GHz directly without any RF mixing and DACs that can generate RF signals up to 7 GHZ when integrated mixing is enabled. The direct RF sampling technique simplified the architecture of the conventional heterodyne base system and has been applied to a variety of scientific instruments from the base-band to the C-band \cite{liu2021characterizing ,henderson2022advanced,liuRA,liuRF,liu2023higher,liuSQ}. The NG-LLRF has been designed and prototyped for C-band LINACs. In \cite{liu2024next,liu:ipac2024-mocn2,liu2025accel,emilio}, the architecture of the prototype system is described, and the results of the NG-LLRF test and performance evaluation are summarized. The RF data converters in the RFSoC have integrated digital up and down conversion modules, enabling the modulation and demodulation to be fully implemented in the digital domain. Therefore, NG-LLRF offers high flexibility in generating arbitrary-shape RF pulses, which can be applied to LINAC stations that require amplitude or phase modulation. In \cite{liu2025high}, we summarize the high-power tests of the NG-LLRF with a C-band test stand and a prototype Cool Copper Collider (C\(^3\)) structure, and cover a selection of test results with different RF pulse shaping modulation schemes. In this paper, more experimental test results with different pulse modulation schemes from the same test stand as described in \cite{liu2025high} will be demonstrated and analyzed.  

\begin{figure}[!htb]
   \centering
   \includegraphics*[width=1\columnwidth]{ 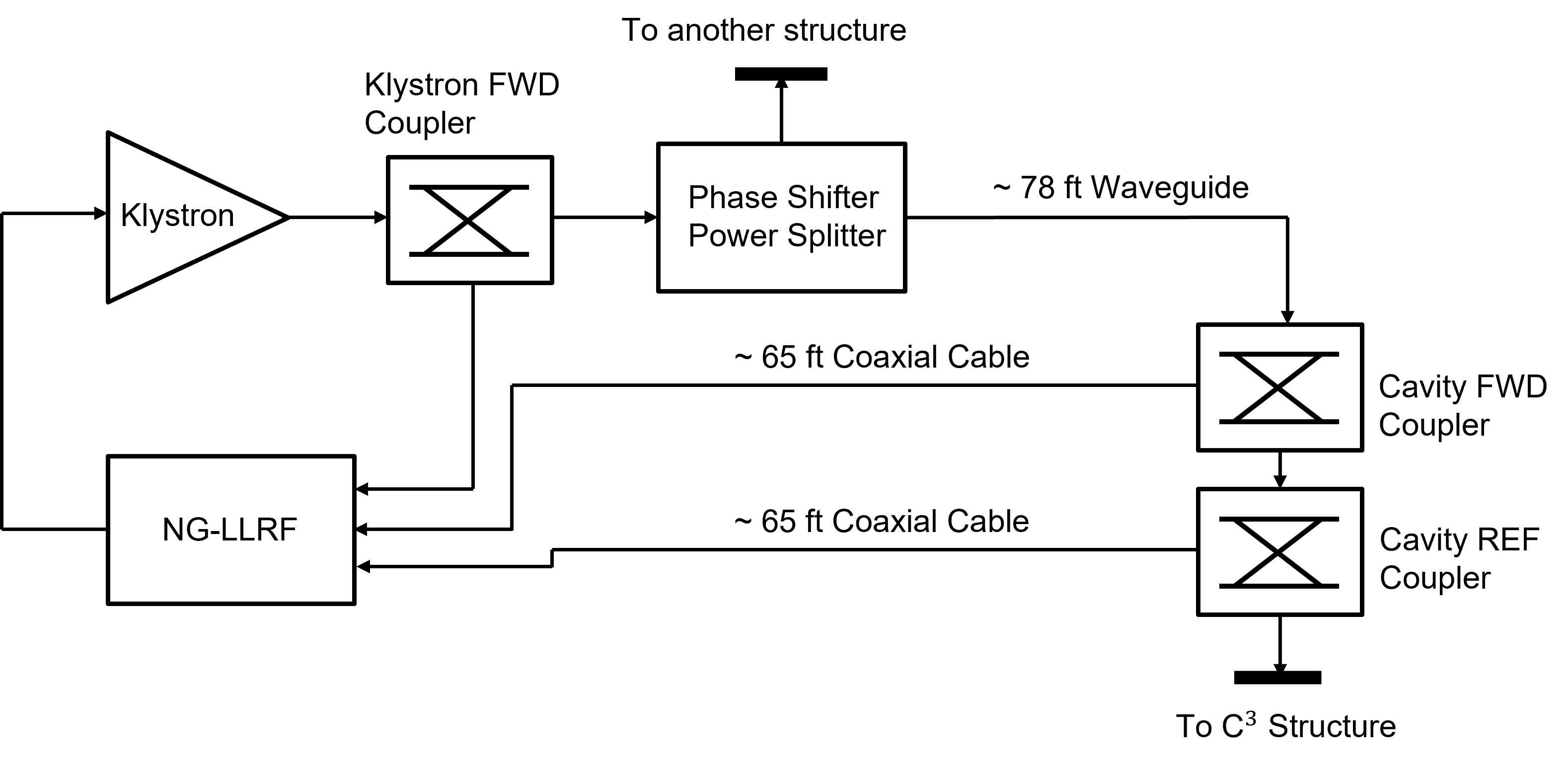}
   \caption{The waveguide schematics of the C-band high-power test setup for the test with different RF pulse shapes \cite{liu2025high}.}
   \label{fig:fig1}
\end{figure}

\section{High-power Test Stand}

\begin{figure}[!htb]
   \centering
   \includegraphics*[width=0.8\columnwidth]{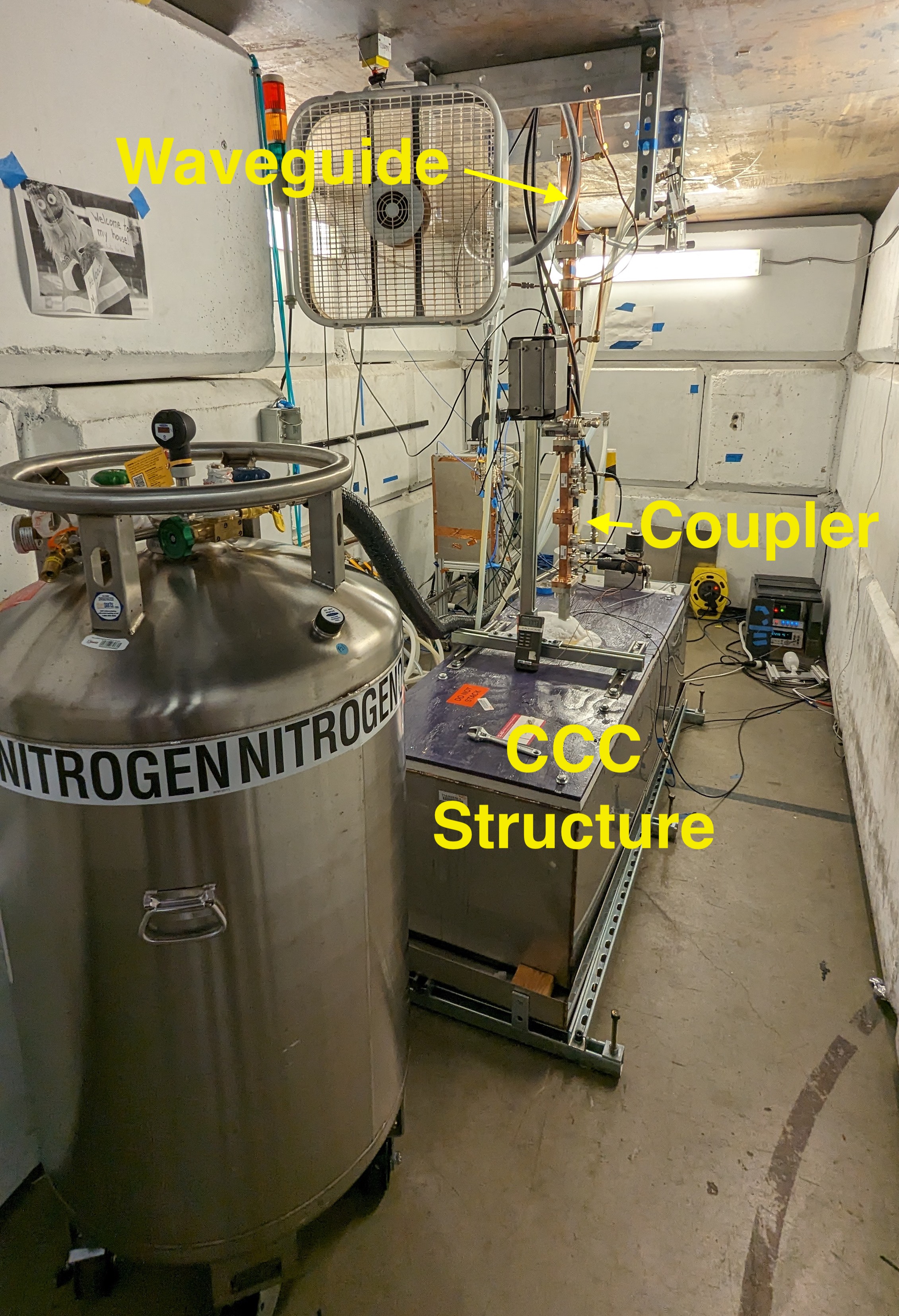}
   \caption{The high-power RF pulse with different shapes injected to a prototype C\(^3\) structure.}
   \label{fig:fig2}
\end{figure}

In the experiment to test different modulation schemes, the full test stand was driven by an NG-LLRF prototype. As Figure \ref{fig:fig1} shows, the RF pulse generated by NG-LLRF is amplified by the klystron and then delivered to the test bunker via a waveguide. The high-power RF is injected into a prototype C\(^3\) in the test bunker shown in Figure \ref{fig:fig2}.  The RF signals from the field sampling couplers at different stages of test stands are looped back and measured by the NG-LLRF. We measured the RF field at three different stages, the klystron forward (FWD), which is directly after the klystron, and the cavity forward (FWD) and cavity reflection (REF) just before the structure as shown in \ref{fig:fig2}. The peak power delivered to the structure ranges from around 4 MW to 16.45 MW for this experiment. The measurement results used in this paper are captured at two peak power levels, 5.17 and 16.45 MW. To avoid damage to the structure, the 5.17 MW peak power is used to operate the system with longer pulses at lower power, and the 16.45 MW peak power is used to operate the system with a shorter pulse and higher power. A more detailed description of the test stand is available in \cite{liu2025high}. 

\section{Pulse Modulation Schemes}

This test aims to demonstrate the high-precision RF pulse generation and measurement capability of NG-LLRF. In this experiment, the test stand was driven by RF pulses modulated with three different schemes, the square pulse, the pulse train, and the pulse with phase reversal, at 10 Hz, with a pulse width of 1 \(\mu\)s, and the peak power of 5.17 MW. The same test procedure was used for all test cases. In software, the baseband waveform is loaded to the BRAM in the FPGA and then the data is accessed by the firmware. The data is interpolated and up mixed with a numerical controlled oscillator around 5.712 GHz by a digital up converter (DUC) in RFSoC. The up-converted samples are streamed to a DAC, which generates the RF pulse. The RF signals from the couplers are attenuated and then captured by the integrated ADCs in the RFSoC. The ADCs samples are down mixed and decimated by a digital down converter (DDC) in RFSoC. The base-band data is recorded in in-phase (I) and quadrature (Q) format. The IQ samples are converted to magnitude and phase for further analysis in this case.

\subsection{Square RF pulse} 

\begin{figure}[!htb]
   \centering
   \includegraphics*[width=1\columnwidth]{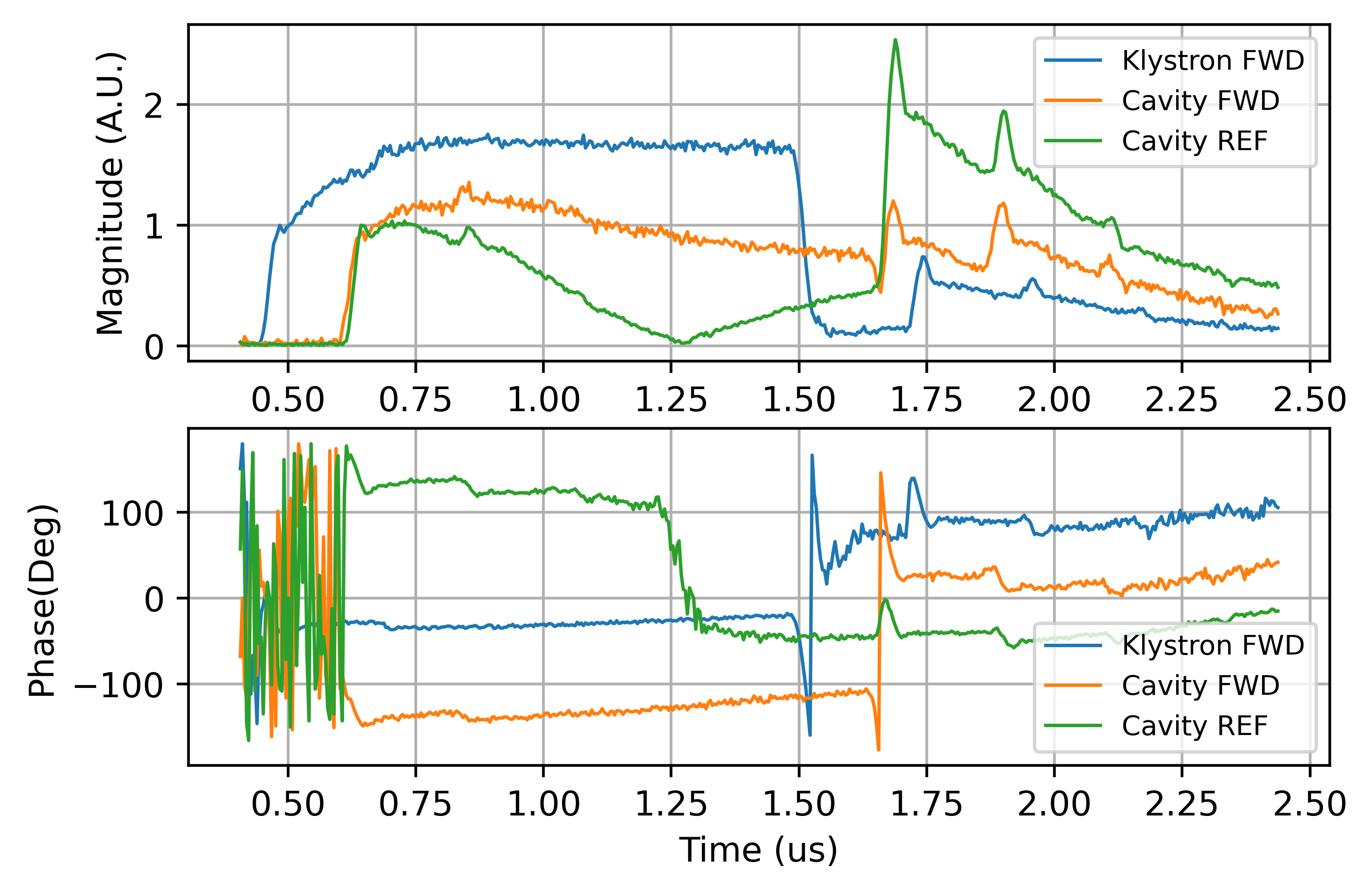}
   \caption{The magnitude and phase of the base-band pulses measured with a 1 \(\mu\)s square RF pulse at the peak power of 5.17 MW.}
   \label{fig:fig3}
\end{figure}

The square RF pulse is the most common waveform used to drive the accelerating structures. Figure \ref{fig:fig3} shows the base-band pulses captured at different stages when the test stand was driven by square RF pulses. The pulses measured with the square pulse drive are used as references for analyzing the pulses measured with other pulse shapes. The magnitude of the klystron forward signal reached a flat top in around 0.3 \(\mu\)s and the flat top is approximately 0.7 \(\mu\) s in this case. The magnitude of the cavity forward signal shows a similar trend with the klystron forward, but with a slight decrease until the RF is off, which can be attributed to cross-coupling with the cavity reflection as observed in \cite{liu2025high}. The magnitude of the cavity reflection decreases linearly with time as the field fills the cavity. The cavity is fully filled in about 0.75 \(\mu\)s, when the magnitude of the reflection signal reaches zero.  The RF power afterwards is fully reflected as the magnitude increases and the phase reverses. When the RF is switched off, there is a sharp rise in the reflection signal, since there is no beam in this case, and the structure was designed to be critically coupled to the beam. Then the magnitude of reflection decays as the field energy dissipates in the structure.

\subsection{Pulse Train} 
\begin{figure}[!htb]
   \centering
   \includegraphics*[width=1\columnwidth]{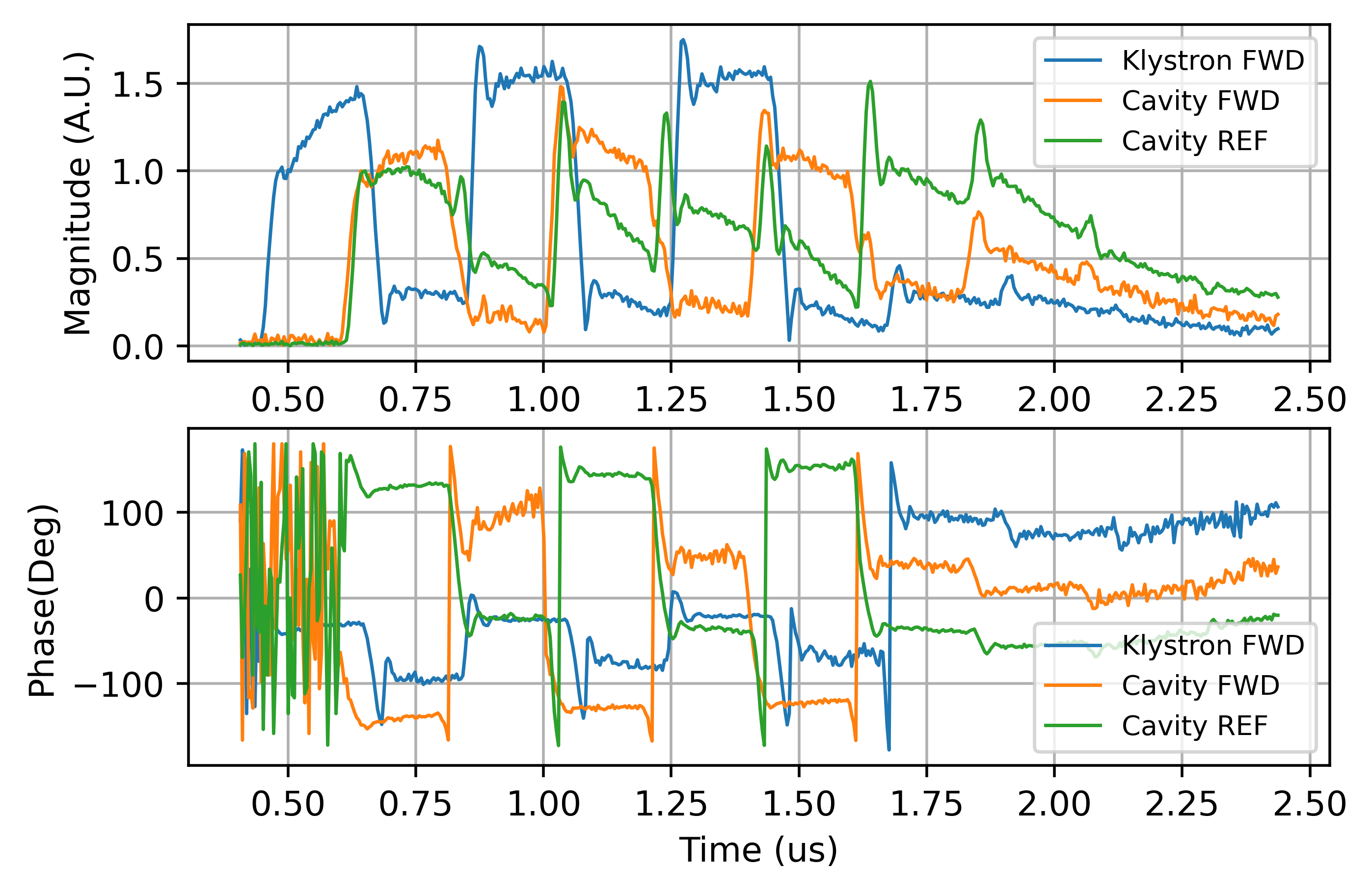}
   \caption{The magnitude and phase of the base-band pulses measured with a 1 \(\mu\)s pulse with RF switch on and off every 0.2 \(\mu\)s at the peak power of 5.17 MW. }
   \label{fig:fig4}
\end{figure}

Figure \ref{fig:fig4} shows the base-band pulses captured at different stages when the test stand was driven by pulses with RF switch on and off every 0.2 \(\mu\)s. As the klystron forward and cavity forward signals show, the desired pulse shape has been produced. The magnitude of the reflection signal declines when the RF is on, but does not reach zero within the pulse, which means that the structure has not been filled.

\subsection{Pulse with Phase Reversal} 
\begin{figure}[!htb]
   \centering
   \includegraphics*[width=1\columnwidth]{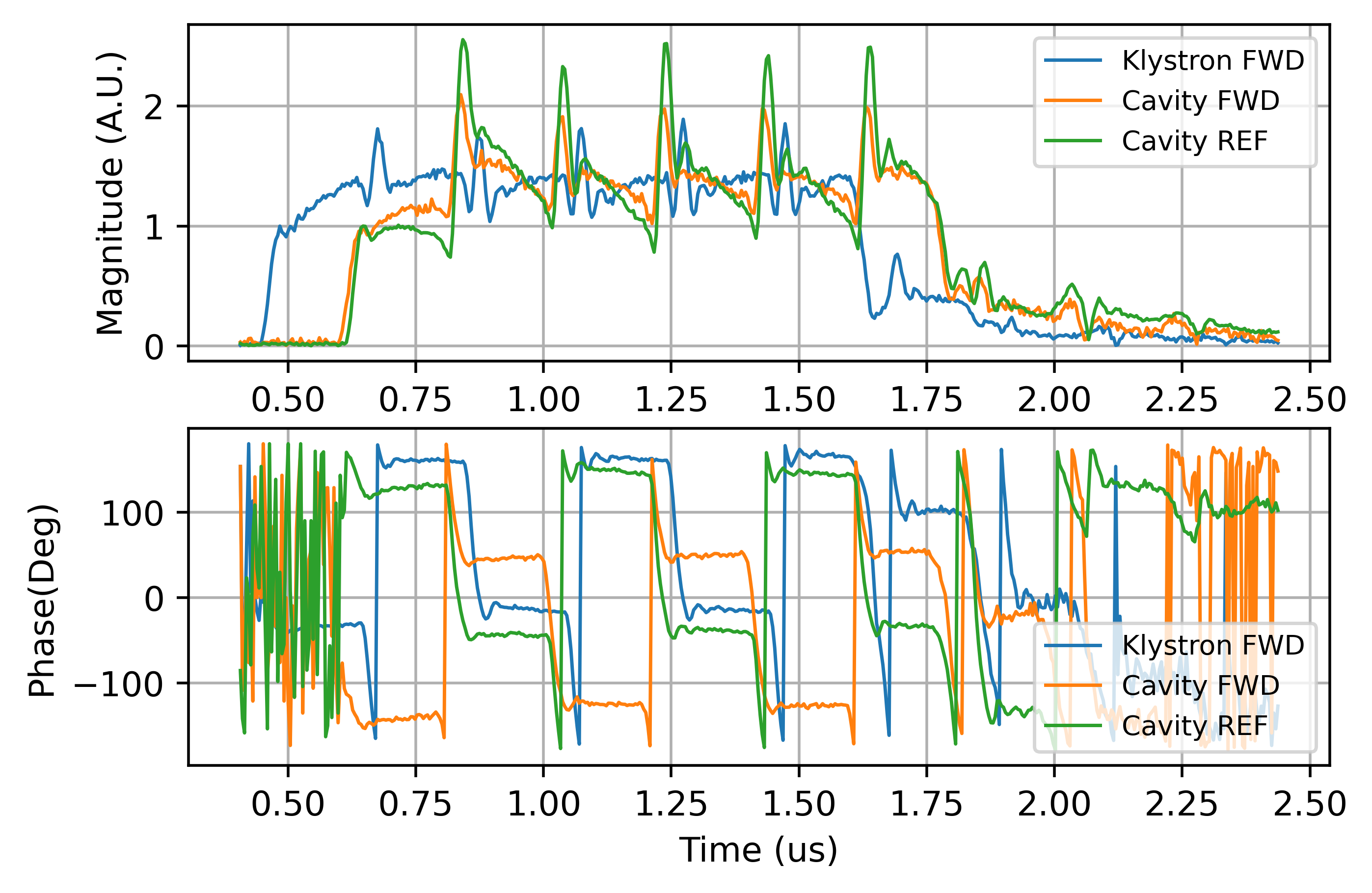}
   \caption{The magnitude and phase of the base-band pulses measured with a 1 \(\mu\)s pulse with RF phase reversal every 0.2 \(\mu\)s at the peak power of 5.17 MW.}
   \label{fig:fig5}
\end{figure}

Figure \ref{fig:fig5} shows the base-band pulses captured at different stages when the test stand was driven by pulses with RF phase reversal every 0.2 \(\mu\)s. The klystron forward and cavity forward signals show that the phase reversals have been successfully introduced. The magnitude and phase of the reflection signal demonstrate that the field energy is extracted when the phase of the drive signal is reversed, and then the RF with the reversed phase starts filling the structure. This cycle repeats until the RF is switched off, and then the reflection signal delays as the energy dissipates in the structure.

\subsection{Pulse with Linear Phase Ramp} 
\begin{figure}[!htb]
   \centering
   \includegraphics*[width=1\columnwidth]{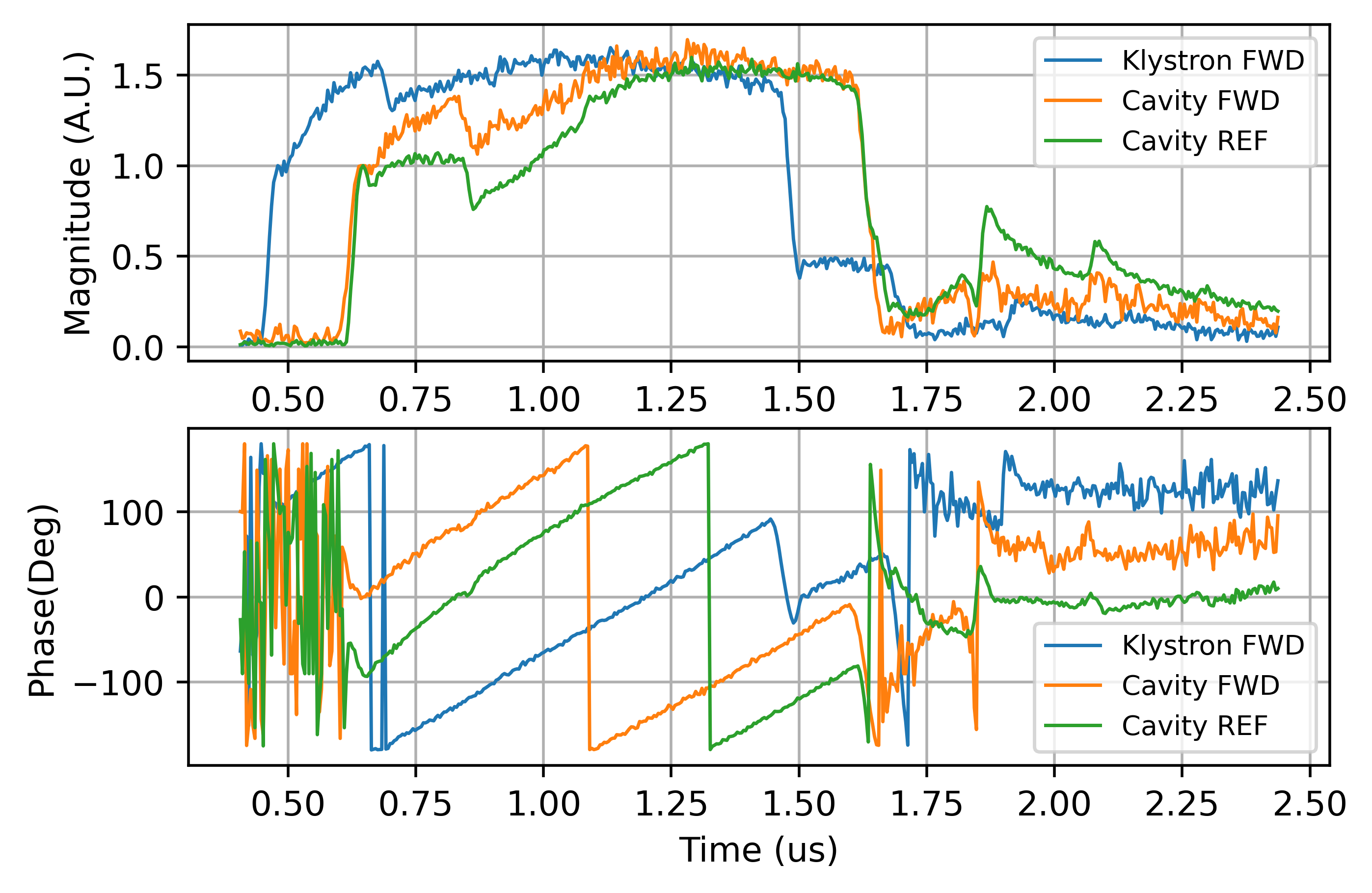}
   \caption{The magnitude and phase of the base-band pulses measured with a 1 \(\mu\)s pulse with a 360\textdegree linear phase ramp at the peak power of 5.17 MW.}
   \label{fig:fig6}
\end{figure}

Figure \ref{fig:fig6} shows the base-band pulses captured at different stages when the test stand was driven by pulses with a linear phase ramp. The klystron forward and cavity forward signals show that the phase ramp has been successfully introduced. However, the RF power injected into the structure is fully reflected, as the phase ramp is equivalent to drive the structure off resonance.

\section{Phase Reversal for SLED}

\begin{figure}[!htb]
   \centering
   \includegraphics*[width=1\columnwidth]{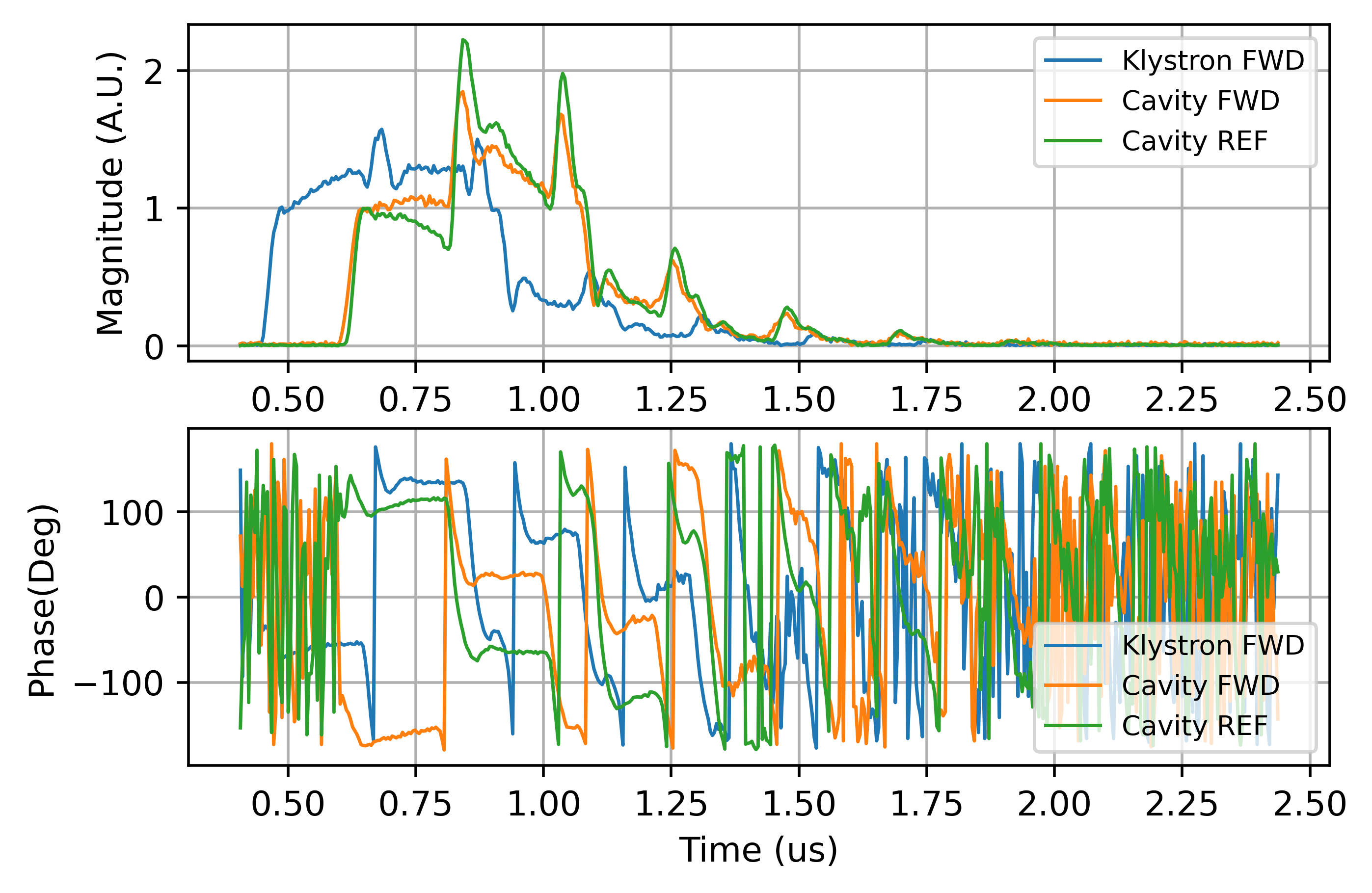}
   \caption{The magnitude and phase of the base-band pulses measured with a 0.45 \(\mu\)s pulse with phase reversal every 0.2 \(\mu\)s at the peak power of 16.45 MW.}
   \label{fig:fig7}
\end{figure}

The phase reversal test has also been performed with a 0.45 \(\mu\)s RF pulse with a peak power of 16.45 MW. The klystron forward and cavity forward signal in Figure \ref{fig:fig7} show that phase reversal has been introduced precisely. The reflection pulse shows that the field fills  the structure for the initial 0.2 \(\mu\)s. When the phase of RF drive is reversed, the magnitude of the reflection signal increased to more than twice the initial peak, and the phase reverses. Rapid power extraction demonstrated the technique required by the SLED.  

\section{Summary}
Direct RF sampling with integrated data converters in RFSoC demonstrated flexibility in generating and measuring arbitrary pulse shapes with the C-band high-power test stand and the prototype C \(^3\) structure. The NG-LLRF system successfully demonstrated the rapid power extraction required by SLED. The duration and relative timing of the phase reversal can be tuned to control a real SLED structure.

%
\ifboolexpr{bool{jacowbiblatex}}%
	{\printbibliography}%
	{%
	
	
} 
%
%


\end{document}